\documentclass[showpacs]{revtex4} 
\usepackage{epsfig,amsmath,amssymb,graphicx,upgreek,textcomp}
\usepackage{natbib}
\usepackage[english]{babel} %

\begin{document}

\vspace{0mm}
\title{The modification of exponents in the Ginzburg-Sobyanin theory of superfluidity} %
\author{Yu.M. Poluektov}
\email{yuripoluektov@kipt.kharkov.ua (y.poluekt52@gmail.com)} %
\affiliation{National Science Center ``Kharkov Institute of Physics and Technology'', 61108 Kharkov, Ukraine} %

\begin{abstract}
A suggested amendment to the temperature dependencies in the
thermodynamic potential of the Ginzburg-Sobyanin theory of
superfluidity, which makes it possible to obtain critical exponents
that are consistent with the general relations of the fluctuation
theory of phase transitions, as well as with modern experimental and
calculated data.
\newline%
{\bf Key words}: %
superfluidity, heat capacity, order parameter, critical exponents,
phase transition temperature, fluctuations
\end{abstract}
\pacs{%
05.30.Jp, 67.25.-k, 67.10.-j, 05.70.Fh, 05.70.Jk }%
\maketitle

\section{Introduction}\vspace{-0mm} 
The theory of helium superfluidity near the critical temperature,
where the quasiparticle description becomes inapplicable, similarly
to the Ginzburg-Landau theory of superconductivity [1], was
developed by Ginzburg and Pitaevsky [2]. A transition to both a
superfluid and a superconducting state is characterized by the
appearance of a complex order parameter $\eta$. However, unlike most
superconductors, with respect to which the theory [1] based on the
mean-field approximation is applied near the phase transition
temperature, the same theory for the transition of liquid helium to
a superfluid state provides only a qualitative description of the
phenomenon. The observed temperature dependencies differ from those
predicted by the mean-field theory. Ginzburg and Sobyanin [3,4]
proposed a modification of the theory [2], in which the temperature
dependence of the coefficients in the expansion of the thermodynamic
potential in powers of the order parameter modulus was selected so
as to obtain dependencies close to those observed. According to the
theory [3,4], the critical heat capacity exponent $\alpha$ is equal
to zero. However, in Ref.\,3, the authors themselves note that there
are experimental indications that this critical exponent is
non-zero. In this case, the theory would have needed to be
clarified, but at that time the authors of Ref.\,3 thought it was
premature to do so.

Currently, critical exponents for superfluid helium are already
measured and calculated with good precision. It can be considered
established that the critical heat capacity exponent is indeed
non-zero. Therefore, the need for some adjustment to the theory has
become apparent.

This study suggests considering two exponents for the dimensionless
temperature in the expansion of the thermodynamic potential near the
$\lambda$-point as phenomenological parameters that should be
determined by comparing them with experimental data. Calculations
show that near the transition temperature of the order parameter,
the behavior of heat capacity, generalized susceptibility, and
correlation length are determined by these two indicators. It turns
out that the known general relations between the critical exponents
hold for arbitrary values of these exponents. The application of
scale invariance makes it possible to establish a relationship
between the dimensionless temperature exponents in the thermodynamic
potential expansion, so that taking this into account the behavior
of all the observed quantities is determined by the only parameter
that can be found from the measured value of some critical exponent.
In this study, this parameter is determined in two ways -- by
measuring the superfluid density and by measuring heat capacity. The
difference in the values obtained is about 0.2\%.

\section{Thermodynamic potential}\vspace{-0mm} %
In the Ginzburg-Sobyanin (GS) theory, the expansion of the
non-equilibrium thermodynamic potential in powers of the complex
order parameter modulus $\eta$ with arbitrary exponents for the
relative temperature $\tau$ can be represented as follows:
\begin{equation} \label{01}
\begin{array}{l}
\displaystyle{%
   \Phi_0=\Phi_I(p,T)-a\tau|\tau|^A|\eta|^2 + \frac{b}{2}|\tau|^B|\eta|^4 + \frac{c}{3}|\eta|^6,   %
}
\end{array}
\end{equation}
where $\tau=(T_\lambda-T)/T_\lambda$, $T_\lambda$ is the superfluid
transition temperature, $\Phi_I(p,T)$ is the thermodynamic potential
of a unit volume of the normal phase near the $\lambda$-point. The
coefficients $a, b, c$ depend on the pressure $p$ but do not depend
on temperature. In the usual formulation of the GS theory [3,4], the
exponents in (1), which, for brevity, will be referred to as
temperature exponents, are as follows: $A_{GS}=1/3$, $B_{GS}=2/3$.
It should also be noted that in (1), the values $A = B = 0$
correspond to the mean-field theory. As shown in Ref.\,3, the fact
that these parameters are different from zero means that the
contribution from fluctuations is effectively taken into account
near the phase transition temperature. We will assume these two
exponents to be non-negative arbitrary parameters of the theory, in
order to find them later by comparing them with the experimental and
calculated data available. If the condition $A+1-2B\geq 0$ is
satisfied, the inclusion in (1) of the last term proportional to
$|\eta|^6$ does not affect the results obtained. Here, the equality
sign is achieved for the exponents in the usual GS theory, and a
more general case is discussed below. Therefore, in Eq.\,(1), we
will further assume the following: $c = 0$ and $a > 0, b > 0$. The
latter conditions ensure the stability of the asymmetric phase.
Under spatially inhomogeneous conditions, it is also necessary to
take into account a contribution from the order parameter gradient
\begin{equation} \label{02}
\begin{array}{l}
\displaystyle{%
   \Phi_g=\frac{\hbar^2}{2m}|\nabla\eta|^2, %
}
\end{array}
\end{equation}
where $m$ is the mass of the helium atom. The effect of some
external field $h$ on a system with a complex order parameter is
normally described [3] as follows:
\begin{equation} \label{03}
\begin{array}{l}
\displaystyle{%
  \Phi_h=-\frac{1}{2}(h\eta^*+h^*\eta).
}%
\end{array}
\end{equation}
The total potential is the sum of all contributions:
$\Phi=\Phi_0+\Phi_g+\Phi_h$. It makes it possible to calculate the
temperature dependences of all observed quantities near the
$\lambda$-point and express the critical exponents in terms of the
parameters $A$ and $B$.

\section{Order parameter, entropy, heat capacity}\vspace{-0mm} %
It follows from the extremum condition of the thermodynamic
potential $\partial\Phi_0/\partial|\eta|^2=0$ that at $\tau>0$ %
\begin{equation} \label{04}
\begin{array}{l}
\displaystyle{%
  |\eta|^2=\frac{a}{b}\,\tau^{A-B+1}\sim\tau^{2\beta}.
}%
\end{array}
\end{equation}
For the critical exponents of the order parameter,this helps to
determine that $2\beta = A - B + 1$. The equilibrium thermodynamic
potential (1) in a superfluid phase at $\tau >0$ takes the following form: %
\begin{equation} \label{05}
\begin{array}{l}
\displaystyle{%
   \Phi_0=\Phi_I(p,T)-\frac{a^2}{2b}\,\tau^{2A-B+2}.    %
}
\end{array}
\end{equation}
Entropy near the transition temperature:
\begin{equation} \label{06}
\begin{array}{l}
\displaystyle{%
   S_S=S_I-\frac{a^2}{2bT_\lambda}(2A-B+2)\,\tau^{2A-B+1},     %
}
\end{array}
\end{equation}
where $S_I = - (\partial\Phi_I/\partial T)_p$ is the entropy in the
normal phase near the transition temperature. The behavior of heat
capacity is described by the following formula:
\begin{equation} \label{07}
\begin{array}{l}
\displaystyle{%
   C_{sp}-C_{Ip}=\frac{a^2}{2bT_\lambda}(2A-B+2)(2A-B+1)\,\tau^{2A-B}\sim\tau^{-\alpha}.  %
}%
\end{array}
\end{equation}
Hence, for the critical heat capacity exponent, we have the
following: $\alpha = B - 2A$. As noted above, according to the usual
formulation of the GS theory, $\alpha_{GS} = B_{GS} - 2A_{GS} =0$.

It should be noted that taking into account the long-wavelength
fluctuations that are not accounted for in the renormalized
thermodynamic potential (1) results in an additional logarithmic
contribution to heat capacity, which is not considered here [3,4].

\section{Generalized susceptibility}\vspace{-0mm} %
In an external field, the effect of which is taken into account in
Eq.\,(3), the equilibrium value of the order parameter is determined
by the following equation:
\begin{equation} \label{08}
\begin{array}{l}
\displaystyle{%
   -a\tau|\tau|^A\eta + b|\tau|^B|\eta|^2\eta - \frac{h}{2}=0.   %
}%
\end{array}
\end{equation}
Let us extract the modulus and phase out of the order parameter and
the external field, and represent them as follows: %
$\eta=|\eta|e^{i\varphi}$ and $h=|h|e^{i\theta}$. Then we can obtain
the following from Eq.\,(8): $\sin(\varphi-\theta) = 0$. Thus, the
following relations between phases are possible: $\varphi=\theta$
and $\varphi=\theta+\pi$. The former case corresponds to the minimum
energy (3). With this in mind, Eq.\,(8) which determines the
dependence of the order parameter modulus on temperature and field
magnitude, takes the following form:
\begin{equation} \label{09}
\begin{array}{l}
\displaystyle{%
   -a\tau|\tau|^A|\eta| + b|\tau|^B|\eta|^3 - \frac{|h|}{2}=0.   %
}%
\end{array}
\end{equation}
Hence, in the normal and superfluid phases, we can determine the
generalized susceptibility in a weak field:
\begin{equation} \label{10}
\begin{array}{ccc}
\displaystyle{%
  \chi\equiv\lim_{|h|\rightarrow
  0}\frac{\partial|\eta|}{\partial|h|}=
   \left\{
     \begin{array}{l}
       \displaystyle{\frac{1}{a|\tau|^{A+1}}, \quad \tau<0, } \vspace{2mm} \\  %
       \displaystyle{\frac{1}{2a\tau^{A+1}}, \quad \tau>0. }
     \end{array} \!\right.
}%
\end{array}
\end{equation}
Thus, according to Eq.\,(10), the critical susceptibility exponent
$\chi\sim|\tau|^{-\gamma}$ is as follows: $\gamma = A +1$. %
Unlike magnetic transitions, in the case of a superfluid system an
external field and susceptibility are largely formal quantities,
since it is unclear how to physically achieve the effect of a
certain field described by the potential (3), and therefore we will
not consider the case of strong fields.

\section{Correlation length}\vspace{-0mm}  %
The calculation of the correlation function for the theory with the
non-equilibrium thermodynamic potential $\Phi=\Phi_0+\Phi_g$ is
given in the Appendix below, with the following formulas obtained
for the normal and superfluid phases:
\begin{equation} \label{11}
\begin{array}{ccc}
\displaystyle{%
  G(r)=
   \left\{
     \begin{array}{l}
       \displaystyle{\frac{Tm}{2\pi\hbar^2r}e^{-r\!/\xi}, \hspace{19mm} \tau<0, } \vspace{2mm} \\  %
       \displaystyle{\frac{Tm}{4\pi\hbar^2r}\left(1+e^{-\sqrt{2}\,r\!/\xi}\right), \quad \tau>0. }
     \end{array} \!\right.
}%
\end{array}
\end{equation}
These formulas include the correlation length, which is determined
as follows:
\begin{equation} \label{12}
\begin{array}{l}
\displaystyle{%
   \xi\equiv\frac{\hbar}{\sqrt{2ma|\tau|^{A+1}}}.   %
}%
\end{array}
\end{equation}
It follows from this that the critical correlation length exponent
$\xi\sim|\tau|^{-\nu}$ is $\nu=\frac{A+1}{2}$. The exponent $\zeta$
is also introduced, which determines the decay of the correlation
function with distance at $\tau=0$: $G(r)\sim r^{-(d-2+\zeta)}$, $d$
being the space dimension [5]. In the three-dimensional case under
consideration, $\zeta=0$.

\section{Critical exponents}\vspace{-2mm} %
In the previous sections, the critical exponents for the
$\lambda$-transition are expressed by two temperature exponents:
\begin{equation} \label{13}
\begin{array}{ccc}
\displaystyle{%
  2\beta=A-B+1, \quad \alpha=B-2A,  %
}\vspace{2mm}\\ %
\displaystyle{\hspace{0mm}%
  \gamma=A+1, \quad \nu=\frac{A+1}{2}, \quad \zeta=0.  %
}%
\end{array}
\end{equation}
From this it is immediately clear that, in the GS theory with
arbitrary exponents $A, B$, the following general relation holds
between the critical exponents:
\begin{equation} \label{14}
\begin{array}{ccc}
\displaystyle{%
  \alpha+2\beta+\gamma=2,  %
}%
\end{array}
\end{equation}
which was first obtained by Essam and Fisher [6,7]. Another general
relation is also satisfied [5]:
\begin{equation} \label{15}
\begin{array}{ccc}
\displaystyle{%
  \nu(2-\zeta)=\gamma.  %
}%
\end{array}
\end{equation}
Moreover, since $\zeta=0$, then $2\nu=\gamma$.

Equations (14) and (15) are not related to any assumptions about the
nature of the fluctuation pattern near the transition temperature.
If we use the hypothesis of scale invariance [5--8], this will lead
to another relation between the critical exponents:
\begin{equation} \label{16}
\begin{array}{ccc}
\displaystyle{%
  \nu d=2-\alpha,  %
}%
\end{array}
\end{equation}
where $d$ is the dimension of space. In the three-dimensional case
under consideration, $3\nu=2 - \alpha$. This, taking into account
(13), provides a relationship between the temperature exponents:
\begin{equation} \label{17}
\begin{array}{ccc}
\displaystyle{%
  B=\frac{(A+1)}{2}.  %
}%
\end{array}
\end{equation}
Thus, the use of the scale invariance hypothesis allows us to
confine ourselves to the only fitting parameter that determines the
temperature dependence of all quantities near the
$\lambda$-transition, and all critical exponents can be expressed by
this parameter, for example:
\begin{equation} \label{18}
\begin{array}{ccc}
\displaystyle{%
  2\beta=\frac{(A+1)}{2}, \quad \alpha=\frac{(1-3A)}{2}, \quad %
  \gamma=A+1, \quad \nu=\frac{A+1}{2}. %
}%
\end{array}
\end{equation}
For the temperature exponent in the initial formulation of the GS
theory [3,4], $A_{GS} = 1/3$, and the corresponding values are as follows: %
$\beta_{GS} = 1/3$, $\alpha_{GS} = 0$, $\gamma_{GS} = 4/3$, $\nu_{GS} = 2/3$. %

To determine the quantity $A$, it is enough to measure one of the
critical exponents (18). The critical exponents for the
$\lambda$-transition in liquid helium ($^4$He) are the most
accurately measured of all critical exponents. In Ref.\,9, the
authors measured the speed of the second sound near the
$\lambda$-point to determine the temperature dependence of
superfluid density $\rho_s\sim|\eta|^2\sim\tau^{2\beta}$, where
\begin{equation} \label{19}
\begin{array}{ccc}
\displaystyle{%
  2\beta=0.6705 \pm 0.0006. %
}%
\end{array}
\end{equation}
The values of the $A$ and $B$ exponents calculated using the
measured value (19) are given in the first row of Table I.

In Refs.\,10--12, the authors measured the critical heat capacity
exponent $\alpha$:
\begin{equation} \label{20}
\begin{array}{ccc}
\displaystyle{%
  \alpha=-0.01285 \pm 0.00038, \qquad [10],  %
}\vspace{2mm}\\ %
\displaystyle{\hspace{0mm}%
  \alpha=-0.01056 \pm 0.00038, \qquad [11],  %
}\vspace{2mm}\\ %
\displaystyle{\hspace{0mm}%
  \alpha=-0.0127\pm 0.0003, \qquad\hspace{4mm} [12].  %
}%
\end{array}
\end{equation}
All measurements were carried out in space in order to avoid the
blurring of the transition due to gravity. The difference in the
obtained values (20) is associated with the ambiguity of
interpretation. To find $A$, the author of this study used the last
value given in (20), which was obtained in Ref.\,12. The values of
the $A$ and $B$ exponents calculated from measurements of the heat
capacity exponent are given in the second row of Table I.
\begin{table}[h!] \nonumber
\vspace{-3mm}%
\caption{The values of $A$ and $B$.} %
\vspace{0.5mm}%
\begin{tabular}{|c|c|c|c|c|c|} \hline 
\rule{3mm}{0pt} $A$ \rule{3mm}{0pt}           & \rule{3mm}{0pt} $B$ \rule{3mm}{0pt} & \rule{3mm}{0pt} $2\beta$ \rule{3mm}{0pt}  & \rule{3mm}{0pt} $\alpha$  \rule{3mm}{0pt} & \rule{3mm}{0pt} $\gamma$ \rule{3mm}{0pt} & \rule{3mm}{0pt} $\nu$ \rule{3mm}{0pt} \\ \hline %
0.3410                                        &  0.6705                             &  $\,0.6705 \pm 0.0006$ [9]                &  $-0.0115$                               &  1.3410                                  &  0.6705                                \\ \hline %
0.3418                                        &  0.6709                             &  0.6709                                   &  $\,-0.0127 \pm 0.0003$ [12]             &  1.3418                                  &  0.6709                                \\ \hline %
\end{tabular}  
\end{table}

\noindent %
The differences between the experimental values and the
corresponding values of the initial theory [3,4] (as shown in Table
I) are determined by the following relations: %
$|A-A_{GS}|/A_{GS}\approx 0.0231$, $|B-B_{GS}|/B_{GS}\approx 0.0057$ %
when determining from the order parameter exponent [9], and %
$|A-A_{GS}|/A_{GS}\approx 0.0255$, $|B-B_{GS}|/B_{GS}\approx 0.0063$ %
when determining from the heat capacity exponent [12]. Thus, the
relative temperature exponent in the second term of the
thermodynamic potential (1) has changed by less than 3\%, %
and in the third term -- by less than 1\%. %

For comparison, let us also give the values of the critical
exponents as calculated in Refs.\,13--15. In Ref.\,13, the authors
used a field theoretic renormalization group method to obtain the
following values:
\begin{equation} \label{21}
\begin{array}{ccc}
\displaystyle{%
  \gamma_{PS}=1.3172 \pm 0.0008, \qquad   %
  \nu_{PS}   =0.6700 \pm 0.0006.          %
}%
\end{array}
\end{equation}
Using lattice methods, the authors of Ref.\,14 produced the
following estimates of the critical exponents for the superfluid
transition in $^4$He:
\begin{equation} \label{22}
\begin{array}{ccc}
\displaystyle{%
  \beta_{CHPV}=0.3486 \pm 0.0001, \qquad \alpha_{CHPV}=-0.0151 \pm 0.0003,  %
}\vspace{2mm}\\ %
\displaystyle{\hspace{0mm}%
  \gamma_{CHPV}=1.3178 \pm 0.0002, \qquad \nu_{CHPV}=0.6717 \pm 0.0001.  %
}%
\end{array}
\end{equation}
In Ref.\,15, the following values were calculated:
\begin{equation} \label{23}
\begin{array}{ccc}
\displaystyle{%
  \beta_{SN}=0.3479 \pm 0.0016, \qquad \alpha_{SN}=-0.0117 \pm 0.0031,  %
}\vspace{2mm}\\ %
\displaystyle{\hspace{0mm}%
  \gamma_{SN}=1.3159 \pm 0.0008, \qquad \nu_{SN}=0.6706 \pm 0.0010.  %
}%
\end{array}
\end{equation}
Thus, both the experiment [9,12] and the calculations [13--15]
clearly indicate that the critical heat capacity exponent $\alpha$
is different from zero.

It should be noted that for the self-consistency of the theory it is
important to satisfy Eq.\,(17), which is a consequence of scale
invariance. In this case, in the thermodynamic potential (1), the
temperature dependence of the term proportional to $|\eta|^6$ which
has been neglected, is the same as that of the terms with $|\eta|^2$
and $|\eta|^4$, therefore this has not affected the obtained
temperature exponents. The relative contribution of possible terms
with $|\eta|^8$, $|\eta|^{10}$ and higher exponents decreases as the
transition temperature is approached, and therefore it is
justifiable to neglect such terms. However, it is worth noting that
in order to ensure the stability of systems with $b < 0$ it should
also be necessary to take the term with $|\eta|^6$ into account in
Eq.\,(1), assuming that $c > 0$.

\section{Conclusion }\vspace{-2mm} %
Accurate measurements [9--12] and calculations [13--15] of critical
exponents near the $\lambda$-transition in $^4$He show that the
critical heat capacity exponent is different from zero, which
contradicts the initial formulation of the Ginzburg-Sobyanin theory
[3,4]. However, as shown in this study, this shortcoming can be
easily corrected if exponents for the dimensionless temperature in
the thermodynamic potential are considered as fitting parameters.
Given the scale invariance hypothesis, all critical exponents in
$^4$He can be expressed by a single exponent for the dimensionless
temperature. When specifying the critical exponent values, it is
necessary for the values of the temperature exponents $A$ and $B$ to
also be specified, though the structure of the GS theory itself will
remain unchanged. It should also be noted that such a formulation of
the theory of phase transitions, which generalizes the mean-field
theory [5], is universal in nature and can be used to describe other
systems, for example such as those that are magnetic.

\appendix
\section{Calculation of the correlation function}\vspace{-2mm} %
The probability of deviation from the equilibrium state is
determined as follows:
\begin{equation} \label{A01}
\begin{array}{ccc}
\displaystyle{%
  w=Ne^{-\Delta\tilde{\Phi}\!/T},  %
}%
\end{array}
\end{equation}
where $N$ is the normalization factor,
$\Delta\tilde{\Phi}=\int\!\Delta\Phi({\bf r})d{\bf r}$ is the
deviation of the thermodynamic potential from the equilibrium value.
Fluctuations are assumed to be small; therefore, we take into
account the deviation from the equilibrium value accurate up to a
quadratic term in the order parameter fluctuations. Since the phase
of the order parameter is not determined in an equilibrium state, we
will not extract the modulus and phase from the complex order
parameter.

{\bf 1}. In the normal phase, where $\tau=-|\tau|<0$:
\begin{equation} \label{A02}
\begin{array}{ccc}
\displaystyle{%
  \Delta\Phi({\bf r})=a|\tau|^{A+1}|\eta|^2 + \frac{\hbar^2}{2m}|\nabla\eta|^2.  %
}%
\end{array}
\end{equation}
Using the expansion
\begin{equation} \label{A03}
\begin{array}{ccc}
\displaystyle{%
  \eta({\bf r})=\frac{1}{\sqrt{V}}\sum_{{\bf k}}\eta_{{\bf k}}e^{i{\bf k}{\bf r}},   %
}%
\end{array}
\end{equation}
we can obtain the following:
\begin{equation} \label{A04}
\begin{array}{ccc}
\displaystyle{%
  \Delta\tilde{\Phi}=\sum_{{\bf k}}L_{{\bf k}}\big(\eta_{{\bf k}}'^2 + \eta_{{\bf k}}''^2\big),   %
}%
\end{array}
\end{equation}
where the real and imaginary parts are highlighted $\eta_{{\bf
k}}=\eta_{{\bf k}}'+i\eta_{{\bf k}}''$, and the following notation
is introduced:
\begin{equation} \label{A05}
\begin{array}{ccc}
\displaystyle{%
  L_{{\bf k}}=a|\tau|^{A+1} + \frac{\hbar^2k^2}{2m}.   %
}%
\end{array}
\end{equation}
Therefore, the normalized probability (A.1) can be written as
\begin{equation} \label{A06}
\begin{array}{ccc}
\displaystyle{%
  w=\prod_{{\bf k}} \frac{L_{{\bf k}}}{\pi T}\,e^{ \,-\frac{L_{{\bf k}}\left(\eta_{{\bf k}}'^2 + \eta_{{\bf k}}''^2\right)}{T} }.  %
}%
\end{array}
\end{equation}
The correlation function is defined by the expression
\begin{equation} \label{A07}
\begin{array}{ccc}
\displaystyle{%
  G(r)=\big\langle\eta({\bf r}_1)\eta^*({\bf r}_2)\big\rangle =  %
  \frac{1}{V}\sum_{\,{\bf k}_1,{\bf k}_2 }\big\langle\eta_{{\bf k}_1}\eta_{{\bf k}_2}^*\big\rangle\,e^{i({\bf k}_1{\bf r}_1-{\bf k}_2{\bf k}_2)} = %
}\vspace{2mm}\\ %
\displaystyle{\hspace{0mm}%
= \frac{1}{V}\sum_{\,{\bf k}_1,{\bf k}_2 }\big[\big\langle\eta_{{\bf k}_1}'\eta_{{\bf k}_2}' + \eta_{{\bf k}_1}''\eta_{{\bf k}_2}''  %
-i\eta_{{\bf k}_1}'\eta_{{\bf k}_2}''+i\eta_{{\bf k}_1}''\eta_{{\bf k}_2}'\big\rangle\big]\,e^{i({\bf k}_1{\bf r}_1-{\bf k}_2{\bf k}_2)}, %
}%
\end{array}
\end{equation}
where $r = |{\bf r}_1-{\bf r}_2|$. Given the expression for the
probability (A.6), we have the following:
\begin{equation} \label{A08}
\begin{array}{ccc}
\displaystyle{\hspace{0mm}%
  \big\langle\eta_{{\bf k}_1}'\eta_{{\bf k}_2}'\big\rangle = \big\langle\eta_{{\bf k}_1}''\eta_{{\bf k}_2}''\big\rangle  %
  =\Delta({\bf k}_1-{\bf k}_2)\frac{T}{2L_{{\bf k}_1}},  \qquad %
  \big\langle\eta_{{\bf k}_1}'\eta_{{\bf k}_2}''\big\rangle = 0. %
}%
\end{array}
\end{equation}
Here $\Delta({\bf k})=1$ if ${\bf k}=0$, and $\Delta({\bf k})=0$ if
${\bf k}\neq 0$. By substituting (A.8) in (A.7), and changing from
summation to integration, we obtain
\begin{equation} \label{A09}
\begin{array}{ccc}
\displaystyle{\hspace{0mm}%
  G(r)=\frac{T}{2\pi^2r}\int_0^\infty\frac{k\sin\!kr}{\left(a|\tau|^{A+1}+\frac{\hbar^2k^2}{2m}\right)}\,dk = %
  \frac{Tm}{\pi^2\hbar^2r}\int_0^\infty\frac{y\sin(k\xi^{-1}y)}{1+y^2}\,dy = %
  \frac{Tm}{2\pi\hbar^2r}\,e^{-r/\xi}, %
}%
\end{array}
\end{equation}
where the correlation length $\xi$ is determined by the following formulas: %
\begin{equation} \label{A10}
\begin{array}{ccc}
\displaystyle{\hspace{0mm}%
  \xi\equiv\xi_0|\tau|^{-\frac{A+1}{2}}, \qquad %
  \xi_0\equiv\frac{\hbar}{\sqrt{2ma}}.          %
}%
\end{array}
\end{equation}
It is easy to verify that the correlation function
\begin{equation} \label{A11}
\begin{array}{ccc}
\displaystyle{\hspace{0mm}%
  \tilde{G}(r)=\big\langle\eta({\bf r}_1)\eta({\bf r}_2)\big\rangle = 0. %
}%
\end{array}
\end{equation}

Having separated the real and imaginary parts $\eta({\bf
r})=\eta'({\bf r})+i\eta''({\bf r})$ from the fluctuation of the
order parameter, we obtain the correlation functions for real
quantities from (A.9) and (A.11):
\begin{equation} \label{A12}
\begin{array}{ccc}
\displaystyle{\hspace{0mm}%
  G'(r)\equiv\big\langle\eta'({\bf r}_1)\eta'({\bf r}_2)\big\rangle =  %
             \big\langle\eta''({\bf r}_1)\eta''({\bf r}_2)\big\rangle =%
\frac{G(r)}{2}=\frac{Tm}{4\pi\hbar^2r}e^{-r\!/\xi},  %
}\vspace{2mm}\\ %
\displaystyle{\hspace{0mm}%
   \tilde{G}'(r)\equiv\big\langle\eta'({\bf r}_1)\eta''({\bf r}_2)\big\rangle = 0.    %
}%
\end{array}
\end{equation}

{\bf 2}. In the superfluid phase, where $\tau>0$:
\begin{equation} \label{A13}
\begin{array}{ccc}
\displaystyle{\hspace{0mm}%
  \eta({\bf r})=\overline{\eta}+\delta\eta({\bf r}), \qquad  %
  \eta^*({\bf r})=\overline{\eta}+\delta\eta^*({\bf r}).  %
}%
\end{array}
\end{equation}
By choosing a phase, the equilibrium value can be made real: %
$\overline{\eta}=\sqrt{a/b}\,\tau^{(A-B+1)\!/2}$. The potential
density fluctuation, accurate up to quadratic terms, is as follows:
\begin{equation} \label{A14}
\begin{array}{ccc}
\displaystyle{\hspace{0mm}%
  \Delta\Phi({\bf r})=\frac{a}{2}\,\tau^{A+1}\big(\delta\eta+\delta\eta^*\big)^2 %
  +\frac{\hbar^2}{2m}\nabla\delta\eta^*\nabla\delta\eta. %
}%
\end{array}
\end{equation}
Using the expansion
\begin{equation} \label{A15}
\begin{array}{ccc}
\displaystyle{\hspace{0mm}%
  \delta\eta({\bf r})=\frac{1}{\sqrt{V}}\sum_{\!{\bf k}}\eta_{{\bf k}}e^{i{\bf k}{\bf r}}, \qquad %
  \delta\eta^*({\bf r})=\frac{1}{\sqrt{V}}\sum_{\!{\bf k}}\eta_{{\bf k}}^*e^{-i{\bf k}{\bf r}},  %

}%
\end{array}
\end{equation}
we obtain the fluctuation of the total potential
\begin{equation} \label{A16}
\begin{array}{ccc}
\displaystyle{\hspace{0mm}%
  \Delta\tilde{\Phi}=\sum_{\!{\bf k}}L_{{\bf k}}|\eta_{{\bf k}}|^2 + %
  \frac{1}{2}\,a\tau^{A+1}\sum_{{\bf k}}\big(\eta_{{\bf k}}^*\eta_{-{\bf k}}^* + \eta_{{\bf k}}\eta_{-{\bf k}}\big). %
}%
\end{array}
\end{equation}
This expression differs from the case of the normal phase (A.4) by
the last term containing the products of amplitudes with opposite
momenta. Now we should move on to new variables using a
transformation similar to the Bogoliubov transformation for operators: %
\begin{equation} \label{A17}
\begin{array}{ccc}
\displaystyle{\hspace{0mm}%
  \eta_{{\bf k}}=u_{{\bf k}}\gamma_{{\bf k}}-\upsilon_{{\bf k}}\gamma_{-{\bf k}}^*, \qquad %
  \eta_{-{\bf k}}^*=u_{{\bf k}}\gamma_{-{\bf k}}^*-\upsilon_{{\bf k}}\gamma_{{\bf k}}. %

}%
\end{array}
\end{equation}
It is assumed, and confirmed by the results, that the coefficients
of this transformation are real and $u_{{\bf k}}=u_{-{\bf k}}$,
$\upsilon_{{\bf k}}=\upsilon_{-{\bf k}}$. By requiring that the
Jacobian of the transition to new variables should be equal to one,
we obtain the following condition for the coefficients:
\begin{equation} \label{A18}
\begin{array}{ccc}
\displaystyle{\hspace{0mm}%
  u_{{\bf k}}^2-\upsilon_{{\bf k}}^2=1. %

}%
\end{array}
\end{equation}
Moving on to to new variables in (A.16), using the transformation
(A.17), and requiring that the products of the variables with
opposite momenta vanish, we arrive at the following equation:
\begin{equation} \label{A19}
\begin{array}{ccc}
\displaystyle{\hspace{0mm}%
  a\tau^{A+1}\big(u_{{\bf k}}^2+\upsilon_{{\bf k}}^2\big) -2u_{{\bf k}}\upsilon_{{\bf k}}L_{{\bf k}}=0. %

}%
\end{array}
\end{equation}
This equation is satisfied if the following conditions are fulfilled: %
\begin{equation} \label{A20}
\begin{array}{ccc}
\displaystyle{\hspace{0mm}%
  L_{{\bf k}}u_{{\bf k}}-a\tau^{A+1}\upsilon_{{\bf k}} = \varepsilon_{{\bf k}}u_{{\bf k}}, %
}\vspace{2mm}\\ %
\displaystyle{\hspace{0mm}%
  -a\tau^{A+1}u_{{\bf k}}+L_{{\bf k}}\upsilon_{{\bf k}} = -\varepsilon_{{\bf k}}\upsilon_{{\bf k}}. %
}%
\end{array}
\end{equation}
The condition that the determinant of this system of homogeneous
linear algebraic equations is equal to zero results in the following equation: %
\begin{equation} \label{A21}
\begin{array}{ccc}
\displaystyle{\hspace{0mm}%
  \varepsilon_{{\bf k}}=\pm\sqrt{L_{{\bf k}}^2-a^2\tau^{2(A+1)}}\,. %
}%
\end{array}
\end{equation}
Taking into account the condition (A.18), we find expressions that
determine the transformation coefficients (A.17):
\begin{equation} \label{A22}
\begin{array}{ccc}
\displaystyle{\hspace{0mm}%
  u_{{\bf k}}^2=\frac{1}{2}\bigg(\frac{L_{{\bf k}}}{\varepsilon_{{\bf k}}}+1\bigg), \qquad%
  \upsilon_{{\bf k}}^2=\frac{1}{2}\bigg(\frac{L_{{\bf k}}}{\varepsilon_{{\bf k}}}-1\bigg), \qquad%
  u_{{\bf k}}\upsilon_{{\bf k}}=\frac{a\tau^{A+1}}{2\varepsilon_{{\bf k}}}. %
}%
\end{array}
\end{equation}
Since $L_{{\bf k}}$ is positive in (A.21), a plus sign should be
chosen. In the new variables, the fluctuation of the total potential
takes the following form:
\begin{equation} \label{A23}
\begin{array}{ccc}
\displaystyle{\hspace{0mm}%
  \Delta\tilde{\Phi}=\sum_{{\bf k}}\varepsilon_{{\bf k}}\big(\gamma_{{\bf k}}'^2+\gamma_{{\bf k}}''^2\big), %
}%
\end{array}
\end{equation}
where $\gamma_{{\bf k}}=\gamma_{{\bf k}}'+i\gamma_{{\bf k}}''$.
Expression (A.23) differs from (A.4) for the normal phase only by
the substitution $L_{{\bf k}}\rightarrow\varepsilon_{{\bf k}}$. Therefore %
\begin{equation} \label{A24}
\begin{array}{ccc}
\displaystyle{\hspace{0mm}%
  \big\langle\gamma_{{\bf k}_1}'\gamma_{{\bf k}_2}'\big\rangle = \big\langle\gamma_{{\bf k}_1}''\gamma_{{\bf k}_2}''\big\rangle  %
  =\Delta({\bf k}_1-{\bf k}_2)\frac{T}{2\varepsilon_{{\bf k}_1}},  \qquad %
  \big\langle\gamma_{{\bf k}_1}'\gamma_{{\bf k}_2}''\big\rangle = 0. %
}%
\end{array}
\end{equation}
For the correlation function, we obtain
\begin{equation} \label{A25}
\begin{array}{ccc}
\displaystyle{\hspace{0mm}%
  G(r)=\big\langle\delta\eta({\bf r}_1)\delta\eta^*({\bf r}_2)\big\rangle =  %
  \frac{T}{V}\sum_{{\bf k}}\frac{L_{{\bf k}}}{\varepsilon_{{\bf k}}^2}e^{i{\bf k}({\bf r}_1-{\bf r}_2)} =  %
}\vspace{2mm}\\ %
\displaystyle{\hspace{0mm}%
  =\frac{Tm}{2\pi^2\hbar^2r}\int_0^\infty\!dk\,\frac{\sin kr}{k}\frac{\Big(a\tau^{A+1}+\frac{\hbar^2k^2}{2m}\Big)}{\Big(a\tau^{A+1}+\frac{\hbar^2k^2}{4m}\Big)} = %
  \frac{Tm}{\pi^2\hbar^2r}\int_0^\infty\!\frac{dy}{y}\sin\big(r\xi^{-1}y\big)\frac{(1+y^2)}{(2+y^2)}.  %
}%
\end{array}
\end{equation}
By calculating the integral, we finally find
\begin{equation} \label{A26}
\begin{array}{ccc}
\displaystyle{\hspace{0mm}%
  G(r)=\frac{Tm}{4\pi\hbar^2r}\Big(1+e^{-\sqrt{2}\,r\!/\xi}\Big). %
}%
\end{array}
\end{equation}
At distances $r\ll\xi$, Eq.\,(A.26) turns to Eq.\,(A.9) for the
normal phase. At $r\gg\xi$, we obtain the following power law: %
$G(r)\sim\frac{Tm}{4\pi\hbar^2r}$.

The ``anomalous'' correlation function is calculated in a similar manner: %
\begin{equation} \label{A27}
\begin{array}{ccc}
\displaystyle{\hspace{0mm}%
  \tilde{G}(r)\equiv\big\langle\delta\eta({\bf r}_1)\delta\eta({\bf r}_2)\big\rangle =  %
  \frac{Tm}{4\pi\hbar^2r}\Big(-1+e^{-\sqrt{2}\,r\!/\xi}\Big).  %
}%
\end{array}
\end{equation}
In contrast to the normal phase (A.11), this function is not equal
to zero in the superfluid phase.

Having separated the real and imaginary parts %
$\delta\eta({\bf r})=\delta\eta'({\bf r})+\delta\eta''({\bf r})$ %
from the fluctuation of the order parameter, we obtain the
correlation functions for real quantities from (A.26) and (A.27):
\begin{equation} \label{A28}
\begin{array}{ccc}
\displaystyle{\hspace{0mm}%
  G'(r)=\big\langle\delta\eta'({\bf r}_1)\delta\eta'({\bf r}_2)\big\rangle =   %
  \frac{1}{2}\big[G(r)+\tilde{G}(r)\big]=\frac{Tm}{4\pi\hbar^2r}\,e^{-\sqrt{2}\,r\!/\xi},  %
}\vspace{2mm}\\ %
\displaystyle{\hspace{0mm}%
  G''(r)=\big\langle\delta\eta''({\bf r}_1)\delta\eta''({\bf r}_2)\big\rangle =   %
  \frac{1}{2}\big[G(r)-\tilde{G}(r)\big]=\frac{Tm}{4\pi\hbar^2r},  %
}\vspace{2mm}\\ %
\displaystyle{\hspace{0mm}%
  \tilde{G}'(r)=\big\langle\delta\eta'({\bf r}_1)\delta\eta''({\bf r}_2)\big\rangle = 0.   %
}%
\end{array}
\end{equation}
One of these, namely $G'(r)$, decays exponentially with distance,
while $G''(r)$ decays according to the power law.



\end{document}